\documentclass[superscriptaddress,showpacs,final,balancelastpage,pra,onecolumn,tightenlines,nofootinbib]{revtex4}
\pdfoutput=1

\usepackage{amsfonts}
\usepackage{amsmath}
\usepackage{amssymb}
\usepackage{graphicx}
\usepackage{times}
\providecommand{\U}[1]{\protect\rule{.1in}{.1in}}
\providecommand{\U}[1]{\protect\rule{.1in}{.1in}}
\providecommand{\U}[1]{\protect\rule{.1in}{.1in}}

\begin{document}

\title{Effective three-body interactions of neutral bosons in optical lattices}

\author{P.R. Johnson}
\email{pjohnson@american.edu}
\affiliation{Department of Physics, American University, Washington, D.C. 20016, USA}

\affiliation{Joint Quantum Institute, National Institute of Standards and Technology and
University of Maryland, Gaithersburg, Maryland 20899, USA}

\author{E. Tiesinga}
\affiliation{Joint Quantum Institute, National Institute of Standards and Technology and
University of Maryland, Gaithersburg, Maryland 20899, USA}

\author{J.V. Porto}
\affiliation{Joint Quantum Institute, National Institute of Standards and Technology and
University of Maryland, Gaithersburg, Maryland 20899, USA}

\author{C.J. Williams}
\affiliation{Joint Quantum Institute, National Institute of Standards and Technology and
University of Maryland, Gaithersburg, Maryland 20899, USA}



\begin{abstract}

We show that there are \emph{effective} three- and higher-body interactions
generated by the two-body collisions of atoms confined in the lowest
vibrational states of a 3D optical lattice. The collapse and revival dynamics
of approximate coherent states loaded into a lattice are a particularly
sensitive probe of these higher-body interactions; the visibility of
interference fringes depend on both two-, three-, and higher-body energy
scales, and these produce an initial dephasing that can help explain the
surprisingly rapid decay of revivals seen in experiments. If inhomogeneities
in the lattice system are sufficiently reduced, longer timescale partial and
nearly full revivals will be visible. Using Feshbach resonances or control of
the lattice potential it is possible to tune the effective higher-body
interactions and simulate effective field theories in optical lattices.

\end{abstract}

\date{November 29, 2009}

\maketitle

\section{Introduction}

The collapse and revival of matter-wave coherence is an expected consequence
of two-body atom-atom interactions in trapped Bose-Einstein condensates (BECs)
\cite{Wright1996,Wright1997,Milburn1997,Walls1997}. Collapse and revival of
few-atom coherent states in optical lattices has been seen in a number of
experiments, first in single-well lattices \cite{Greiner2002 Collapse and
Revival} and subsequently in double-well lattices
\cite{Anderlini2006,JSebbyStrabley1}. In these experiments, a BEC is quickly
loaded into a fairly deep 3D lattice such that the quantum state approximately
factors into a product of coherent states localized to each lattice site
\cite{Pethick2002,Haroche2006,Bloch2008}. Each coherent state, which is a
superposition of different atom-number states, initially has a well-defined
phase. If the lattice potential is quickly turned off before atom-atom
interactions have a significant influence, the coherent states released from
confinement at each site expand and overlap resulting in interference fringes
in the imaged atom-density. However, if the atoms are held in the lattice for
a longer duration before release, interactions will play a significant role by
causing the phases of the different atom-number states in the superposition at
each site to evolve at different rates. This will result in a dephasing of the
coherent state, and a subsequent collapse of the interference fringe
visibility after the atoms are released. For atoms in a homogenous lattice
with two-body interactions and negligible tunneling, the coherent states at
each lattice site are predicted to revive when the atom-number component
states simultaneously re-phase after multiples of the time $t_{2}=2\pi
\hbar/U_{2}$, where $U_{2}$ is the two-body interaction energy
\cite{Wright1996,Wright1997,Milburn1997,Walls1997,Greiner2002 Collapse and
Revival}.

In addition to the expected two-body physics described above, we show that the
data in \cite{Greiner2002 Collapse and Revival,Anderlini2006,JSebbyStrabley1}
should also contain strong signatures of\emph{ coherent} three- and
higher-body interactions. In contrast to the coherent dynamics described in
this paper, recent experiments have studied inelastic three-body processes,
including recent observations of Efimov physics
\cite{Efimov1970,Kraemer2006,Bedaque2000}, by tracking atom loss from
recombination \cite{Burt1997}. There has been a growing interest in three- and
four-body physics (e.g.,
\cite{Esry1999,Petrov2005,Stoll2005,Braaten2006,Ferlaino2009,Huckans2009}),
and the role of intrinsic three-body interactions on equilibrium quantum
phases in optical lattices has been studied in
\cite{Chen2008,Schmidt2008,Sansone2009}. The influence of higher bands on the
Mott-insulator phase transition has been analyzed in \cite{Lutchyn2009}, and
three-body interactions of fermions and polar molecules in lattices have also
been explored \cite{Buchler2007}.

In this paper, we use the ideas of effective field theory to show that virtual
transitions to higher vibrational states generate \emph{effective,
}coherent\emph{ }three-body interactions between atoms in the lowest
vibrational states of a deep 3D lattice where tunneling can be neglected. More
generally, virtual excitations also generate effective four- and higher-body
interactions giving the non-equilibrium dynamics multiple energy scales. We
also show that loading coherent states into an optical lattice creates a
sensitive interferometer for probing higher-body interactions. In a
sufficiently uniform lattice, multiple frequencies manifested as beatings in
the visibility of the collapse and revival oscillations give a direct method
for measuring the energy and frequency scales for elastic higher-body
interactions. Remarkably, multiple-frequency collapse and revival patterns
have been seen in recent experiments \cite{DAMOP2009}.

Three-body interactions can also explain the surprisingly rapid damping of
revivals seen in \cite{Greiner2002 Collapse and
Revival,Anderlini2006,JSebbyStrabley1}, where the overall visibility of the
interference fringes decays after roughly 5 revivals ($\sim3$ ms for the
system parameters in \cite{Greiner2002 Collapse and
Revival,Anderlini2006,JSebbyStrabley1}). This short timescale cannot be
explained in terms of tunneling or atom loss. For example, for the system
parameters in \cite{Greiner2002 Collapse and
Revival,Anderlini2006,JSebbyStrabley1}, the tunneling-induced decoherence
timescale has been found to be a factor of 10-100 times too long
\cite{Fischer2008 Tunneling induced decay}, and the atom loss from three-body
recombination \cite{Burt1997} appears to be negligible \cite{DAMOP2009}. The
latter observation is consistent with the expected three-body recombination
timescales for $^{87}$Rb in a lattice \cite{3Boby recombo footnote}.

The damping of revivals can be partially explained by the expected variation
in $U_{2}$ over a non-uniform lattice due to an additional harmonic term in
the trapping potential. Inhomogeneity in $U_{2}$ causes dephasing due to the
variation in the revival times for coherent states at different sites,
however, the estimated 3-5\% inhomogeneity of $U_{2}$ should allow as many as
10-20 revivals compared to the $\sim$5 seen in \cite{Greiner2002 Collapse and
Revival,Anderlini2006,JSebbyStrabley1}. In contrast, we show below that
coherent three-body interactions can cause dephasing of coherent states at
each lattice site after only a few revivals.

The effective theory in this paper describes the low-energy, small scattering
length, small atom number per lattice site regime, for deep 3D lattices with
negligible tunneling. These approximations are reasonable for the experiments
in \cite{Greiner2002 Collapse and Revival,Anderlini2006,JSebbyStrabley1}.
Extensions of the analysis might include tunneling, including second-order
\cite{Folling2007} and interaction driven \cite{Ananikian2006} tunneling, and
the incorporation of intrinsic higher-body interactions. Effective field
theory has also proven to be an important tool in the large scattering length
limit \cite{Bedaque2000}. It would be particularly interesting to simulate the
controlled breakdown of the effective theory developed here by increasing the
scattering length or atom number, or by tuning other lattice parameters.
Looking beyond the realm of atomic physics, our analysis suggests interesting
possibilities for using optical lattices to test important mechanisms in
effective field theory \cite{Effective field theory}.

In Section 2, we construct a multimode Hamiltonian $\hat{H}$ which we use to
obtain an effective single-mode Hamiltonian $\tilde{H}_{\text{eff}}$. In
Section 3, we describe the physical processes that generate higher-body
interactions. In Section 4, we estimate the effective three-body energy. In
Section 5, we show how the coherent three-body interactions modify the
collapse and revival dynamics. Finally, we summarize our results in Section 6.

\section{Effective three-body model for neutral bosons in an optical lattice}

A many-body Hamiltonian for mass $m_{a}$ neutral bosons in a single spin state
can be written as
\begin{align}
\mathcal{H}  &  =\int\hat{\psi}^{\dagger}H_{0}\hat{\psi}d\mathbf{r}+\frac
{1}{2}\int\hat{\psi}^{\dagger}\left(  \mathbf{r}\right)  \hat{\psi}^{\dagger
}\left(  \mathbf{r}^{\prime}\right)  V_{2}\left(  \mathbf{r,r}^{\prime
}\right)  \hat{\psi}\left(  \mathbf{r}\right)  \hat{\psi}\left(
\mathbf{r}^{\prime}\right)  d\mathbf{r}d\mathbf{r}^{\prime}%
\label{Many body Hamiltonian}\\
&  +\frac{1}{6}\int\hat{\psi}^{\dagger}\left(  \mathbf{r}\right)  \hat{\psi
}^{\dagger}\left(  \mathbf{r}^{\prime}\right)  \hat{\psi}^{\dagger}\left(
\mathbf{r}^{\prime\prime}\right)  V_{3}\left(  \mathbf{r,r}^{\prime
}\mathbf{,r}^{\prime\prime}\right)  \hat{\psi}\left(  \mathbf{r}\right)
\hat{\psi}\left(  \mathbf{r}^{\prime}\right)  \hat{\psi}\left(  \mathbf{r}%
^{\prime\prime}\right)  d\mathbf{r}d\mathbf{\mathbf{r}}^{\prime}%
d\mathbf{\mathbf{r}}^{\prime\prime}+...\mathbf{,}\nonumber
\end{align}
where $V_{m}$ are intrinsic $m$-body interaction potentials, and $H_{0}$ is
the Hamiltonian for a single particle in the optical lattice. We set
$V_{m>2}=0$ to focus on the physics of effective interactions induced by
$V_{2}$. In experiments, the effect of intrinsic and effective interactions
are both present.

It is our goal to construct a low energy, effective Hamiltonian $\tilde
{H}_{\text{eff}}$ for describing a small number of atoms in the vibrational
ground state of a lattice site, while incorporating leading-order corrections
from virtual excitation to higher bands. In the quantum mechanical approach,
Huang \emph{et al.} \cite{Huang1957} have shown that a local regularized
delta-function potential $V_{2}\left(  \mathbf{r,r}^{\prime}\right)
\propto\delta^{\left(  3\right)  }\left(  \mathbf{r-r}^{\prime}\right)
\left(  d/d\varrho\right)  \varrho,$ where $\varrho=\left\vert \mathbf{r}%
-\mathbf{r}^{\prime}\right\vert ,$ can be used to obtain the low-energy
scattering for two particles. To go beyond the two-particle case, we find it
convenient to instead use the renormalization methods of quantum field theory
and the non-regularized delta-function potential
\begin{equation}
V_{2}\left(  \mathbf{r,r}^{\prime}\right)  =g_{2}\delta^{\left(  3\right)
}\left(  \mathbf{r-r}^{\prime}\right)  . \label{Delta function potential}%
\end{equation}
We regularize the theory in perturbation theory by using a high-energy cutoff
$\Lambda$ in the sum over intermediate states, which is equivalent to using a
regularized (non-singular) potential. We view $\Lambda$ as a physical
threshold beyond which the low-energy theory fails. We note that the
low-energy physics does not, in the end, depend on the method of
regularization, and that the physical results found below after
renormalization are insensitive to $\Lambda$. The key observation is that even
if a fully regularized form of $V_{2}$ is used renormalization is still
required recognizing that the \emph{bare }parameter $g_{2}$ is not\emph{ }the
physical (renormalized) coupling strength $\tilde{g}_{2}$. (In the following
we use a tilda to distinguish between bare and renormalized parameters.)

Employing renormalized perturbation theory \cite{Effective field theory}, we
write $g_{2}=\tilde{g}_{2}+c,$ where
\begin{equation}
\tilde{g}_{2}=\frac{4\pi\hbar^{2}a_{\text{scat}}}{m_{a}}+\mathcal{O}\left(
a_{\text{scat}}^{2}\right)  , \label{Renormalization condition}%
\end{equation}
is chosen to reproduce the exact, low-energy limit given in \cite{Busch1998}
for two atoms in a spherically symmetric harmonic trap, and $a_{\text{scat}}$
is the scattering length at zero-collisional energy. The first-order
approximation to $\tilde{g}_{2}$ suffices for the calculation of the
three-body energy at second order given below. The value of the counter-term
$c,$ which cancels the contributions to the two-body interaction energy that
diverge with $\Lambda,$ is determined by the normalization condition Eq.
(\ref{Renormalization condition}). The local Hamiltonian with counter-term and
physical coupling parameter becomes
\begin{equation}
\mathcal{H}=\int\hat{\psi}^{\dagger}H_{0}\hat{\psi}d\mathbf{r}+\frac{1}%
{2}\left(  \tilde{g}_{2}+c\right)  \int\hat{\psi}^{\dagger}\hat{\psi}%
^{\dagger}\hat{\psi}\hat{\psi}d\mathbf{r}.
\end{equation}

To develop a low-energy effective theory for a deep optical lattice, we expand
the field over a set of bosonic annihilation operators $\hat{a}_{i\mu}$ and
single particle wavefunctions $\phi_{i\mu}\left(  \mathbf{r}\right)  $ giving
$\hat{\psi}\left(  \mathbf{r}\right)  =\sum_{i\mu}\phi_{i\mu}\left(
\mathbf{r}\right)  \hat{a}_{i\mu},$ where the indices $\mu=\left\{  \mu
_{x},\mu_{y},\mu_{z}\right\}  $ with $\mu_{x,y,z}=0,1,2,...$ label 3D
vibrational states and $i$ labels the lattice sites. To focus on the role of
interactions we assume a deep lattice with $n_{s}\gtrsim3$ states per spatial
dimension at each site, making tunneling of atoms in the ground vibrational
state $\mu=\left\{  0,0,0\right\}  \equiv0$ negligible on the timescale of
interest \cite{Greiner2002 Collapse and Revival}. Since we are not considering
the role of tunneling, for simplicity we use isotropic harmonic oscillator
wavefunctions at each site with frequency $\omega$ and length scale
$\sigma=\sqrt{\hbar/m_{a}\omega}$ determined by the (approximately) harmonic
confinement within a single lattice well. Note that even with tunneling
neglected, anharmonicity of the lattice potential is a potentially significant
effect. We also expect our model to break down or to require significant
modification for very shallow lattices or near the Mott-insulator phase
transition where the effects of tunneling are important \cite{Fisher1989 Bose
Hubbard model,Jaksch1998,Greiner2002 Mott Insulator,Lutchyn2009}.

Inserting the expansion for $\hat{\psi}$ into $\mathcal{H}$, interchanging the
order of integration over $\mathbf{r}$ and summation over modes, and dropping
terms that transfer atoms between sites (e.g. tunneling), we obtain for each
lattice site the multimode Hamiltonian $\hat{H}=\hat{H}_{0}+\hat{H}_{2},$
where
\begin{equation}
\hat{H}_{0}=\sum\nolimits_{\mu}E_{\mu}\hat{a}_{\mu}^{\dagger}\hat{a}_{\mu}
\label{H0}%
\end{equation}
and
\begin{equation}
\hat{H}_{2}=\frac{1}{2}\left(  \tilde{U}_{2}+A\right)  \sum\nolimits_{\mu
\nu\sigma\lambda}K_{\mu\nu\sigma\lambda}\hat{a}_{\mu}^{\dagger}\hat{a}_{\nu
}^{\dagger}\hat{a}_{\sigma}\hat{a}_{\lambda}. \label{Multimode Hamiltonian}%
\end{equation}
For brevity we suppress the lattice site index $i$. The\emph{ }single particle
energies are $E_{\mu}=\left(  \mu_{x}+\mu_{y}+\mu_{z}\right)  \hbar\omega$,
setting the ground state energy $E_{0}\equiv E_{\left\{  0,0,0\right\}  }=0.$
The two-body interaction energy for ground state atoms is%
\begin{equation}
\tilde{U}_{2}=\frac{\tilde{g}_{p}}{\left(  2\pi\right)  ^{3/2}\sigma^{3}%
}=\sqrt{\frac{2}{\pi}}\hbar\omega\left(  a_{\text{scat}}/\sigma\right)  ,
\label{Energy renormalization condition}%
\end{equation}
and $A=\left(  2\pi\right)  ^{-3/2}c/\sigma^{3}\ $is the counter-term in units
of energy. The matrix elements
\begin{equation}
K_{\mu\nu\gamma\delta}=\left(  2\pi\right)  ^{3/2}\sigma^{3}\int\phi_{\mu}%
\phi_{\nu}\phi_{\gamma}\phi_{\delta}d\mathbf{r}
\label{K approx matrix elements}%
\end{equation}
are normalized so that $K_{0000}=1,$ and they vanish for transitions that do
not conserve parity. It should be noted that, when there is a cutoff in the
sum over modes, both the regularized and non-regularized delta-function
potential lead to the same Hamiltonian $\hat{H}$ and matrix elements in
Eq.(\ref{K approx matrix elements}), and thus they produce the same results in
the regularized (cutoff) quantum field theory. We emphasize that after the
renormalization of the two-body interaction energy, the induced three-body
interaction energy is insensitive to the cutoff $\Lambda.$ We develop the
perturbation theory in the small parameter $\xi$ defined by%
\begin{equation}
\xi\equiv\frac{\tilde{U}_{2}}{\hbar\omega}=\sqrt{\frac{2}{\pi}}\frac
{a_{\text{scat}}}{\sigma}+\mathcal{O}\left(  a_{\text{scat}}^{2}\right)  .
\end{equation}

The\emph{ }total interaction energy for $n$ atoms in the vibrational ground
state in the single mode per site approximation is $E_{\text{int}}=\tilde
{U}_{2}n\left(  n-1\right)  /2.$ Commonly, a single-mode approximation is made
based on the two-body interaction energy \emph{per particle }being much less
than the band gap, i.e., $E_{\text{int}}/n=\tilde{U}_{2}\left(  n-1\right)
/2\ll\hbar\omega$ or $n\xi\ll1.$ For $^{87}$Rb with scattering length
$a_{\text{scat}}\simeq5.3$ nm and a lattice with $\omega/2\pi\simeq30$ kHz, we
have $\tilde{U}_{2}/h\simeq2.0$ kHz, and $\xi=0.07.$ We will use these as
typical system parameters in the following analysis. With $\xi=0.07,$ the
single mode per site condition $n\ll\xi^{-1}\sim15\ $is easily satisfied and
the influence of higher-bands will produce only small (though important)
corrections. For coherent states, for example, we show that small three-body
energies can lead to large phase shifts over time resulting in
interferometric-like sensitivity to higher-body and higher-band processes.

To obtain an effective Hamiltonian $\tilde{H}_{\text{eff}},$ we use the
multi-mode Hamiltonian $\hat{H}=\hat{H}_{0}+\hat{H}_{2}$ to compute the
atom-number dependent energy shift for atoms in the vibrational ground state.
Our approach is essentially equivalent to the effective field theory procedure
of summing up to a cutoff over all `high-energy' modes $\mu$ with $E_{\mu}%
\geq\hbar\omega,$ which generates a low energy effective theory with all
consistent local interactions. We obtain an effective Hamiltonian $\tilde
{H}_{\text{eff}}$ for the $\mu=0$ mode that is valid in the low-energy regime
$E_{\text{int}}/n\sim n\tilde{U}_{2}\ll\hbar\omega,$ which is consistent with
the single mode approximation discussed above. Of course the multimode
Hamiltonian $\hat{H}$ itself is an effective Hamiltonian which is only valid
for energy scales $E_{\mu}+E_{\text{int}}/n\ll\hbar/(m_{a}a_{\text{scat}}%
^{2}).$

In the case of atoms confined in a deep well, the effective Hamiltonian for
$\tilde{U}_{2}\ll\hbar\omega$ is
\begin{equation}
\tilde{H}_{\text{eff}}=E_{0}\hat{a}^{\dagger}\hat{a}+\sum_{m>1}\tilde{U}%
_{m}\hat{a}^{\dagger m}\hat{a}^{m}/m!,
\end{equation}
where $\hat{a}^{\dagger}$ creates an atom in a renormalized ground vibrational
state. The $E_{0}\hat{a}^{\dagger}\hat{a}$ term vanishes since we set
$E_{0}=0$. The dominant term in $\tilde{H}_{\text{eff}}$ is the two-body
energy, and the higher-body interaction energies scale as $n\tilde{U}%
_{m}/\tilde{U}_{m-1}\sim\left(  n\tilde{U}_{2}/\hbar\omega\right)  \sim
n\xi\ll1$.

The energies $\tilde{U}_{m}$ can be computed in perturbation theory in the
small parameter $\xi$ using $\hat{H}$ to find the energy of $n$ atoms in the
ground vibrational mode. At $m^{th}$ order in $\xi,$ all local interactions up
through the $\left(  m+1\right)  $-body term $\tilde{H}_{m+1}=\tilde{U}%
_{m+1}\hat{a}^{\dagger m+1}\hat{a}^{m+1}/\left(  m+1\right)  !$ are generated.
In this paper, we work to second order in $\xi$ for which the effective
Hamiltonian is
\begin{equation}
\tilde{H}_{\text{eff}}=\tilde{U}_{2}\hat{a}^{\dagger2}\hat{a}^{2}/2+\tilde
{U}_{3}\hat{a}^{\dagger3}\hat{a}^{3}/6. \label{Effective Hamiltonian}%
\end{equation}
Using $\hat{n}=\hat{a}^{\dagger}\hat{a}$ and $\left[  \hat{a},\hat{a}%
^{\dagger}\right]  =1,$ the two- and three-body terms can be written as
$\hat{a}^{\dagger2}\hat{a}^{2}=\hat{n}\left(  \hat{n}-1\right)  $ and $\hat
{a}^{\dagger3}\hat{a}^{3}=\hat{n}\left(  \hat{n}-1\right)  \left(  \hat
{n}-2\right)  ;$ the latter expression shows explicitly that the effective
three-body interaction only arises when there are three or more atoms in a
well. Eigenstates of $\tilde{H}_{\text{eff}}$ with $n$ atoms have energies
\begin{equation}
\tilde{E}\left(  n\right)  =\tilde{U}_{2}n\left(  n-1\right)  /2+\tilde{U}%
_{3}n\left(  n-1\right)  \left(  n-2\right)  /6. \label{Effective energy}%
\end{equation}
Note that the three-body energy scales as $n^{3}$ and thus its influence
relative to the two-body term, though small, can be tuned by changing the
number of atoms in a well.

\section{Mechanism for effective interactions}%

\begin{figure}
[ptb]
\begin{center}
\includegraphics[
height=1.6853in,
width=6.4056in
]%
{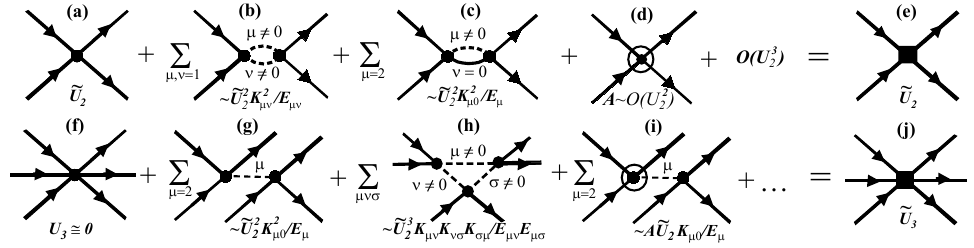}%
\caption{The effective two-body interaction energy $\tilde{U}_{2}$ is given
through second order by diagrams (a)-(d). Diagram (d) is the counter-term that
cancels the diagrams (b) and (c), fixing $\tilde{U}_{2}$ as the physical
(renormalized) two-body energy. Diagrams (f)-(i) are examples of processes
contributing to the effective three-body interaction energy $\tilde{U}_{3},$
represented by diagram (j). Diagram (g) gives the leading order contribution,
assuming $U_{3}=0;$ it shows how an effective three-body interaction involving
three distinct incoming particles arises at second order in perturbation
theory. Diagrams (h) and (i) are two of the effective three-body processes
that arise at third order (others are not shown). If the bare three-body
vertex shown in (f) does not vanish additional three (and higher) body
counter-terms are also required.}%
\end{center}
\end{figure}

We now describe the virtual processes that give rise to effective $m$-body
interactions in a deep lattice. Writing the perturbative expansion for the
energy of an $n$ atom state $\left\vert n\right\rangle $ through second order
as $\tilde{E}\left(  n\right)  =E^{\left(  0\right)  }\left(  n\right)
+E^{\left(  1\right)  }\left(  n\right)  +E^{\left(  2\right)  }\left(
n\right)  ,$ the zeroth-order energy is $E^{\left(  0\right)  }\left(
n\right)  =E_{0}n=0,$ recalling that $E_{0}=0$. The first-order energy shift,
treating $\hat{H}_{2}$ as the perturbation Hamiltonian, is the usual
expression
\begin{equation}
E^{\left(  1\right)  }\left(  n\right)  =\left\langle n\right\vert \hat{H}%
_{2}\left\vert n\right\rangle =\tilde{U}_{2}n\left(  n-1\right)  /2.
\label{1st order energy shift}%
\end{equation}
This is the leading order result for the two-body interaction energy and,
setting $n=2,$ the renormalization condition Eq.
(\ref{Energy renormalization condition}) shows that $A=0$ to first order in
$\tilde{U}_{2}$. Figure~1(a) represents this first-order process.

The second-order energy shift can be written as%
\begin{equation}
E^{\left(  2\right)  }\left(  n\right)  =-\frac{\tilde{U}_{2}^{2}}{4}\sum
_{\mu\geq\nu}^{\Lambda}\frac{s_{\mu\nu}K_{\mu\nu}^{2}\left\vert \left\langle
\mu\nu\right\vert \hat{a}_{\mu}^{\dagger}\hat{a}_{\nu}^{\dagger}\hat{a}%
_{0}\hat{a}_{0}\left\vert n\right\rangle \right\vert ^{2}}{E_{\mu\nu}%
}+An\left(  n-1\right)  /2 \label{2nd order energy shift}%
\end{equation}
with $K_{\mu\nu}\equiv K_{\mu\nu00}$ and $\mu\geq\nu.$ The $\mathcal{O}\left(
\tilde{U}_{2}^{2}\right)  $ counter-term $A$ now appears. At this order, $A$
is determined by the renormalization condition $\tilde{E}\left(  2\right)
=E^{\left(  0\right)  }\left(  2\right)  +E^{\left(  1\right)  }\left(
2\right)  +E^{\left(  2\right)  }\left(  2\right)  =\tilde{U}_{2},$ implying
that $E^{\left(  2\right)  }\left(  2\right)  =0.$ The sum is over
intermediate states $\left\vert \mu\nu\right\rangle \equiv\hat{a}_{\mu
}^{\dagger}\hat{a}_{\nu}^{\dagger}\hat{a}_{0}\hat{a}_{0}\left\vert
n\right\rangle $ with energy $E_{\mu\nu}=E_{\mu}+E_{\nu}>0;$ this excludes the
$\mu=\nu=0$ state. For regularization purposes we introduce a high-energy
cutoff that limits the sum to $E_{\mu\nu}\leq\Lambda$. The factor $s_{\mu\nu
}=\left\{  4,1\right\}  $ if $\left\{  \mu=\nu,\mu\neq\nu\right\}  $ comes
from the two equivalent terms $\hat{a}_{\mu}^{\dagger}\hat{a}_{\nu}^{\dagger}$
and $a_{\nu}^{\dagger}\hat{a}_{\mu}^{\dagger}$ that appear in $\hat{H}_{2}.$
Each term in $E^{\left(  2\right)  }$ involves a two-body collision-induced
transition to a virtual intermediate state. For example, the state $\left\vert
1_{x}1_{x}\right\rangle $ corresponds to two atoms both excited along the $x$
direction with energy $E_{11}=2\hbar\omega$ (note that $K_{11}^{2}=1/4$ for
this transition), with the remaining $n-2$ atoms left in the $\mu=0$ mode.
Because collisions conserve parity, contributions from states like $\left\vert
1_{x}1_{y}\right\rangle $ vanish.

The crucial observation is that the series in Eq.
(\ref{2nd order energy shift}) separates into two distinct sums corresponding
to two-body and three-body interactions, respectively, i.e.,
\begin{equation}
E^{\left(  2\right)  }\left(  n\right)  =\delta U_{2}n\left(  n-1\right)
/2+\delta U_{3}n\left(  n-1\right)  \left(  n-2\right)  /6,
\end{equation}
where $\delta U_{2}$ includes the counter-term contribution $A$ from Eq.
(\ref{2nd order energy shift}). For $\mu\neq0$ \emph{and} $\nu\neq0$
intermediate states, $\left\vert \left\langle \mu\nu\right\vert \hat{a}_{\mu
}^{\dagger}\hat{a}_{\nu}^{\dagger}\hat{a}_{0}\hat{a}_{0}\left\vert
n\right\rangle \right\vert ^{2}=\kappa n\left(  n-1\right)  $ where
$\kappa=\left\{  2,1\right\}  $ if $\left\{  \mu=\nu,\mu\neq\nu\right\}  ,$
with the factor of $2$ resulting from Bose stimulation when both atoms
transition to the same excited state. Because these terms are proportional to
$n\left(  n-1\right)  $ they contribute to the two-body energy shift $\delta
U_{2}$. A diagram representing this two-body process, with two atoms
colliding, making transitions to virtual excited vibrational states, and then
returning to the ground state after a second collision with each other, is
shown in Fig.~1(b). The $\mu\neq0$ virtual states and $\mu=0$ vibrational
ground states are represented by dashed and solid lines, respectively.

The origin of the three-body energy can be seen by examining the $\mu>0,\nu=0$
intermediate states. We have%
\begin{equation}
|\left\langle \mu\nu\right\vert \hat{a}_{\mu}^{\dagger}\hat{a}_{0}^{\dagger
}\hat{a}_{0}\hat{a}_{0}\left\vert n\right\rangle |^{2}=n\left(  n-1\right)
^{2}=n\left(  n-1\right)  +n\left(  n-1\right)  \left(  n-2\right)  ,
\label{Three body matrix element}%
\end{equation}
showing that these terms generate both effective two- and three-body energies.
The extra factor of $\left(  n-1\right)  $ in Eq.
(\ref{Three body matrix element}) results from Bose stimulation of an atom
back into the $\mu=0$ state when two atoms collide but only one makes a
transition to an excited state. Figure~1(c) shows the two-body process
corresponding to the $n\left(  n-1\right)  $ term in Eq.
(\ref{Three body matrix element}). Figure~1(d) shows the counter-term $A$
whose value is determined such that it cancels the contributions from
Figs.~1(b) and (c), thereby maintaining, through second order, the
renormalization condition that the parameter $\tilde{U}_{2}$ is equal to the
physical two-body energy. To arbitrary order the renormalization condition
determines $A$ such that all higher-order two-body diagrams cancel, as
represented by Fig.~1(e).

Figure~1(g) shows the effective three-body process corresponding to the
$n\left(  n-1\right)  \left(  n-2\right)  $ term in Eq.
(\ref{Three body matrix element}). This process gives the leading-order
contribution to $\delta U_{3}$ and generates a three-body interaction energy
$\tilde{U}_{3}=U_{3}+\delta U_{3}$ even if the bare $U_{3},$ represented by
Fig.~1(f), vanishes. More generally, we expect $U_{3}\neq0,$ but nevertheless
the contribution to $\tilde{U}_{3}$ given by $\delta U_{3}$ can be a
significant (possibly even dominant) correction. Looking at Fig.~1(g), we see
that two initial $\mu=0$ atoms collide giving rise to one $\mu\neq0$ atom that
subsequently collides with a \emph{third}, distinct $\mu=0$ atom. In Fig.~1(g)
there are three distinct incoming atoms resulting in an effective three-body
interaction mediated by the $\mu\neq0$ intermediate state. The renormalized
three-body interaction energy is represented in Fig.~1(j) by a square vertex
with three incoming and outgoing particles. Figures~1(h) and (i) show examples
of two different processes contributing to $\tilde{U}_{3}$ at third-order in
$\xi;$ they illustrate how higher-order processes, including counter-terms,
arise. Their contributions, and other third-order processes not shown, are not
explicitly computed below. At third order, effective four-body interactions
also arise.

Notice that there are two types of diagrams in Fig.~1: tree diagrams [e.g.
Fig.~1(g)] and loop diagrams [e.g. Fig.~1(b)]. In general in quantum field
theory the contributions from some loop diagrams diverge with the cutoff
$\Lambda,$ necessitating the need for renormalization, whereas the
contributions from tree diagrams are finite \cite{Effective field theory}. We
will see this behavior explicitly below. In fact, at $m^{th}$ order in $\xi,$
there will be a set of tree diagrams giving a \emph{finite}, leading-order
contribution to the effective $\left(  m+1\right)  $-body interaction energies
$\tilde{U}_{m+1}$. We note that even if all intrinsic higher-body interactions
exactly vanish there will be effective $m$-body interactions and associated
energy scales $\tilde{U}_{m}$ generated by the two-body interactions.
Consequently, the nonequilibrium dynamics of $n$ atoms in the ground
vibrational mode, when $n\xi\ll1,$ will be characterized by a hierarchy of
frequencies $\left(  \tilde{U}_{2}/h,\tilde{U}_{3}/h...,\tilde{U}%
_{m}/h\right)  .$

\section{Estimate of the effective three-body interaction energy}

Returning to Eq. (\ref{2nd order energy shift}) for the second-order energy
shift and separating it into two- and three-body parts, we find that%
\begin{equation}
\delta U_{2}=-\tilde{U}_{2}^{2}\left(  \sum\nolimits_{\mu,\nu}^{\Lambda}%
K_{\mu\nu}^{2}/E_{\mu\nu}\right)  +A,
\end{equation}
and%
\begin{equation}
\delta U_{3}=-6\tilde{U}_{2}^{2}\left(  \sum\nolimits_{\mu>0}^{\Lambda}%
K_{\mu0}^{2}/E_{\mu0}\right)  .
\end{equation}
In the expression for $\delta U_{2}$ the sum is over all $\mu$ and $\nu$ (both
$\mu>\nu$ and $\nu>\mu$) except for the $\mu=\nu=0$ mode. Similarly, in the
expression for $\delta U_{3}$ all $\mu$ except for $\mu=0$ are summed over.

As expected, the sum $\sum_{\mu,\nu}^{\Lambda}K_{\mu\nu}^{2}/E_{\mu\nu}$
corresponding to the second order, 1-loop diagram in Fig.~1(b) diverges with
$\Lambda,$ reflecting the divergent relationship between the bare $U_{2}$ and
renormalized $\tilde{U}_{2}$ energy parameters. In fact, the sum scales with
the cutoff as $\Lambda^{1/2}$. The renormalization condition that $\tilde
{E}\left(  2\right)  =\tilde{U}_{2}$ determines $A$ by requiring that $\delta
U_{2}=0.$ To second-order, the interaction energy of $n$ atoms is thus%
\begin{equation}
\tilde{E}\left(  n\right)  =\tilde{U}_{2}n\left(  n-1\right)  /2+\delta
U_{3}n\left(  n-1\right)  \left(  n-2\right)  /6,
\end{equation}
assuming $\tilde{U}_{3}=\delta U_{3}.$

After cancelling the two-body corrections with $A,$ the remaining second-order
term gives an induced three-body energy that is insensitive to $\Lambda:$ the
quantity $\sum_{\mu>0}^{\Lambda}K_{\mu0}^{2}/E_{\mu0}$ corresponding to the
second-order tree diagram in Fig.~1(g) converges. Writing
\begin{equation}
\delta U_{3}/\hbar\omega=-\beta\xi^{2},
\end{equation}
this sum can be solved analytically for a spherically symmetric harmonic trap
in the $\Lambda\rightarrow\infty$ limit, and we find \cite{Calculation of U3}
\begin{equation}
\beta=4\sqrt{3}-6+6\log\left(  \frac{4}{2+\sqrt{3}}\right)  \simeq1.34....
\end{equation}
Cutting off the sum at $E_{\mu\nu}/\hbar\omega\leq\Lambda/\hbar\omega=4$
already gives $\beta\simeq1.30$ showing the rapid convergence of the series.
The convergence of this sum is an example of the generic behavior that
contributions from tree diagrams are finite. If the bare $U_{3}$ is zero or
sufficiently small, the effective three-body energy is negative, giving
attractive three-body interactions, and reducing the total interaction energy
for both positive or negative $\tilde{U}_{2}.$

We expect significant corrections due to the anharmonicity of the true lattice
potential. The single-particle energies of higher vibrational states are
lowered on the order of the recoil energy $E_{R},$ defined as the gain in
kinetic energy for an atom at rest that emits a lattice photon. This leads to
a decrease of the energy denominator in Eq. (\ref{2nd order energy shift})
and, for the typical system parameters considered here, this can give an
estimated correction to $\tilde{U}_{3}$ of $10\%$ or more. The matrix elements
$K_{\mu\nu}$ will also have corrections. These effects can be computed
numerically using single-particle band theory.

We have defined our perturbation theory around the zero-collisional energy
limit, but in a trap the collision energy of ground state atoms is on the
order of $\hbar\omega.$ As shown in \cite{Bolda2002}, an improved treatment
replaces the zero-energy scattering length by an effective scattering length
defined as%
\begin{equation}
-\frac{1}{a_{eff}}=-\frac{1}{a_{\text{scat}}}+\frac{1}{2}r_{e}k^{2},
\label{Effective scattering length formula}%
\end{equation}
where the effective range $r_{e}$ is on the order of the van der Waals length
scale $\left(  m_{a}C_{6}/\hbar^{2}\right)  ^{1/4}$ away from a Feshbach
resonance, and the collision energy is $\hbar^{2}k^{2}/m_{a}$ \cite{Gao1998}.
For $^{87}$Rb the van der Waals length is approximately $8$ nm. In a trap the
ground vibrational state wavevector $k\simeq\sigma^{-1}$ produces a fractional
increase in scattering length on the order of $\left(  r_{e}/\sigma\right)
\xi.$ By incorporating the effective scattering length model we can extend the
range of validity of our model.

Even neglecting these corrections, the perturbation theory generated by Eqs.
(\ref{Delta function potential}) and (\ref{Multimode Hamiltonian}) does not
predict the two-body energy $\tilde{U}_{2}$ but instead uses the measured
value, or the exact result calculated by other methods such as Busch \emph{et
al }\cite{Busch1998}, as input from which $\delta U_{3}$ is obtained.
Similarly, the effective theory does not yield the intrinsic three-body
interaction energy $U_{3},$ and therefore $\tilde{U}_{3}=U_{3}+\delta U_{3}$
must also be determined by either measurement or a theory of three-body
physics if the intrinsic interaction energies $U_{m>2}$ are non-zero. On the
other hand, the effective theory shows that even if $U_{m>2}=0$ there are
significant induced three- and higher-body interactions, and if non-zero
$\tilde{U}_{m>2}$ are measured the effective contribution from two-body
processes must be taken into account before the intrinsic higher-body coupling
strengths can be extracted. Note that if non-zero bare (intrinsic) parameters
$U_{m>2}$ are included in our model, additional counter-terms will be needed
to cancel divergences, reflecting the need to ultimately determine any
intrinsic higher-body coupling strengths via either measurement or an exact
high-energy theory.

Assuming $U_{3}\simeq0,$ Fig.~2 shows $\tilde{U}_{2}=\xi\hbar\omega$ and
$\tilde{U}_{3}$ versus $\xi,$ including positive ($\xi>0$) and negative
($\xi<0$) scattering lengths. Using $\xi=0.07$ for $^{87}$Rb in a $30$ kHz
well gives $\tilde{U}_{2}/h\simeq1.9$ kHz and $\tilde{U}_{3}/h\simeq-200$ Hz.
Using a Feshbach resonance \cite{Tiesinga1993} to change $a_{\text{scat}}$ and
thus $\xi,$ or fixing $a_{\text{scat }}$ and changing the trap frequency
$\omega,$ it is possible to tune the relative strengths of the three-body (and
higher-body) interactions. It would be interesting to explore the breakdown of
the perturbative model by increasing either $\xi$ or the atom number $n,$ or
by decreasing the lattice depth so that the influence of tunneling and
higher-band effects increases.%

\begin{figure}
[ptb]
\begin{center}
\includegraphics[
height=2.3698in,
width=4.343in
]%
{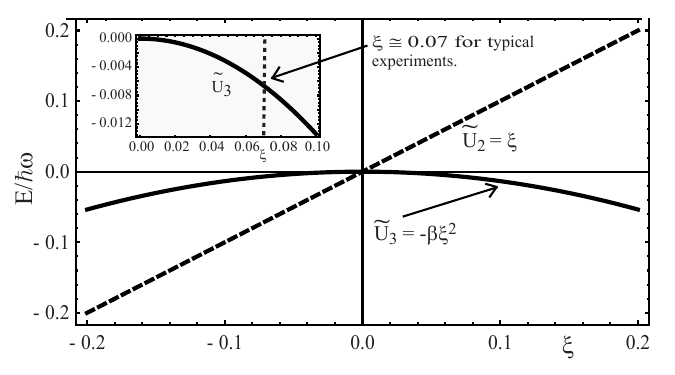}%
\caption{The figure shows $\tilde{U}_{3}$ and $\tilde{U}_{2},$ in units of
$\hbar\omega,$ versus $\xi.$ The bold line shows the induced three-body energy
$\delta\tilde{U}_{3}=\tilde{U}_{3}=-\beta\xi^{2}$ with $\beta=1.34,$ assuming
the intrinsic energy $U_{3}$ vanishes. The dashed line shows the leading-order
two-body energy $\tilde{U}_{2}.$ The graphs extends beyond the regime of
strict validity of the perturbation theory, which requires $\xi n\ll1$ where
$n$ is the number of atoms in a lattice well, to illustrate the overall
scaling of the two- and three-body energies. The collapse and revival
experiments in \cite{Greiner2002 Collapse and
Revival,Anderlini2006,JSebbyStrabley1} have $\omega/2\pi\sim30$ kHz and
$\xi\sim0.07,$ putting them well within the perturbative regime. The inset
shows $\tilde{U}_{3}$ for the range $0<\xi<0.1$. }%
\end{center}
\end{figure}

\section{Dynamics and decoherence of atom-number coherent states}

We now investigate the influence of effective three-body interactions on the
phase coherence of an $N$ atom nonequilibrium state $\left\vert \Psi\left(
0\right)  \right\rangle =(\sum_{i=1}^{M}\hat{a}_{i0}^{\dagger}\left\vert
0\right\rangle /\sqrt{M})^{\otimes N}$, obtained by quickly loading a BEC into
a lattice with $M$ sites. To a good approximation the state can be treated as
the product of coherent states \cite{Greiner2002 Collapse and
Revival,Bloch2008},
\begin{equation}
\left\vert \Psi\left(  0\right)  \right\rangle \simeq\prod_{i}\exp(\sqrt
{\bar{n}_{i}}\hat{a}_{i}^{\dagger})\left\vert 0\right\rangle \simeq\prod
_{i}\left\vert \alpha_{i}\right\rangle ,\label{Coherent states}%
\end{equation}
where $\hat{a}_{i}\left\vert \alpha_{i}\right\rangle =\alpha_{i}\left\vert
\alpha_{i}\right\rangle $ and $\left\vert \alpha_{i}\right\vert ^{2}=\bar
{n}_{i}$ is the average number of atoms in the $i^{th}$ site. A relative phase
$\phi_{ij}$ between sites $i\neq j$ exists when $\langle\hat{a}_{i}^{\dagger
}\hat{a}_{j}\rangle=\eta e^{i\phi_{ij}}$ and $\eta\neq0$. The initial state
$\left\vert \Psi\left(  0\right)  \right\rangle $ has $\eta=\bar{n}$, and
there are well-defined relative phases ($\phi_{ij}=0$ for all $i,j$ in this
case). In contrast, the equilibrium Mott insulator state, achieved by much
slower loading \cite{Jaksch1998,Greiner2002 Mott Insulator}, has approximate
number states in each well giving $\eta\approx0,$ though there can be some
degree of short-range phase coherence \cite{Roberts2003,Roth2003,Gerbier2005}%
.

\begin{figure}
[ptb]
\begin{center}
\includegraphics[
height=1.7333in,
width=4.938in
]%
{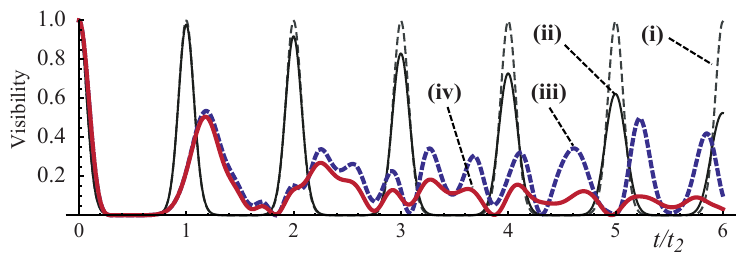}%
\caption{Collapse and revival visibility versus time $t,$ with $\xi=0.07$ and
$\bar{n}=2.5.$ Curve (i) shows the case with neither inhomogeneities nor
three-body interactions included. Curve (ii) shows the effects of $\sim5\%$
inhomogeneities in $U_{2}$. Curve (iii) shows the effects of three-body
interactions with $\beta=1.34$ but no inhomogeneities. Note that the
three-body mechanism influences the visibility of revivals immediately, and it
will be important even if inhomogeneities are stronger than are shown in curve
(ii). Curve (iv) shows the combined effects of inhomogeneities and three-body
interactions.}%
\end{center}
\end{figure}

Coherent states in optical lattices make natural probes of higher-body
coherent dynamics because small atom-number dependent energies can lead to
significant phase shifts over time. After a hold time $t_{h}$ in the lattice,
the initial state evolves to $\left\vert \Psi\left(  t_{h}\right)
\right\rangle \simeq\prod_{i}\left\vert \eta\left(  t_{h}\right)
\right\rangle _{i},$ where the state of the $i^{th}$ well is
\begin{equation}
\left\vert \eta\left(  t_{h}\right)  \right\rangle _{i}=e^{-\bar{n}_{i}/2}%
\sum_{n}\frac{\alpha_{i}^{n}}{\sqrt{n!}}\left\vert n\right\rangle
_{i}e^{-i\tilde{E}_{i}\left(  n\right)  t_{h}/\hbar},
\label{Coherent state dynamics}%
\end{equation}
and $\tilde{E}_{i}\left(  n_{i}\right)  $ is given in Eq.
(\ref{Effective energy}), restoring the index $i$ labeling the lattice site.
Snapping the lattice off at time $t_{h}$, the wavefunctions from each well
freely expand for a time $t_{e}$ until they fully overlap, analogous to the
diffraction of light through a many-slit grating.

For a uniform lattice, the fringe visibility is \cite{Greiner2002 Collapse and
Revival}
\begin{equation}
V\left(  t_{h}\right)  =\left\vert \left\langle \eta\left(  t_{h}\right)
\right\vert \hat{a}\left\vert \eta\left(  t_{h}\right)  \right\rangle
\right\vert ^{2}/\bar{n}. \label{Fringe visibility}%
\end{equation}
With no inhomogeneities and setting $\tilde{U}_{3}=0$, we obtain $V\left(
t_{h}\right)  =e^{-2\bar{n}\left[  1-\cos\left(  \tilde{U}_{2}t_{h}%
/\hbar\right)  \right]  }.$ The visibility for $\bar{n}=2.5$ is plotted as the
thin dashed line labeled (i) in Fig.~3, showing the well-known collapse and
revival dynamics with period $t_{2}=h/\tilde{U}_{2}.$ For the $^{87}$Rb system
parameters used here $t_{2}=0.52$ ms.

The thin line labeled (ii) in Fig.~(3) shows the influence of an approximate
5\% variation in the two-body energy $U_{2}$. We average the $a_{i}\left(
t\right)  $ over a 60 lattice-site diameter spherical distribution. While the
effect of inhomogeneities are important, a larger variation in $U_{2}$ then
expected would be required to explain the decay of interference fringes after
only 5 revivals as seen in experiments \cite{Greiner2002 Collapse and
Revival,Anderlini2006,JSebbyStrabley1}. We note that the longer timescale for
three-body recombination can be distinguished from the coherent, number
conserving interactions derived here by tracking changes in total atom number,
and this appears to be negligible on the revival damping timescale
\cite{DAMOP2009}. Other mechanisms, such as non-adiabatic loading
\cite{Hecker-Denschlag2002,Rey2006} and collisions during expansion
\cite{Gerbier2008} will reduce the initial fringe visibility but do not
explain the rapid decay of the visibility versus hold time $t_{h}$.

To compute the visibility with three-body interactions we numerically
evaluate
\begin{equation}
\left\langle \eta\left(  t_{h}\right)  \right\vert \hat{a}\left\vert
\eta\left(  t_{h}\right)  \right\rangle =\alpha e^{-\bar{n}}\sum_{n=0}%
\frac{\bar{n}^{n}}{n!}e^{-in\left[  \tilde{U}_{2}+\tilde{U}_{3}\left(
n-1\right)  /2\right]  t_{h}/\hbar}.
\label{3-body collapse and revival visibility}%
\end{equation}
The bold (blue) dashed line labeled (iii) in Fig.\ 3 shows the visibility
$V\left(  t_{h}\right)  =\left\vert \left\langle \eta\left(  t_{h}\right)
\right\vert \hat{a}\left\vert \eta\left(  t_{h}\right)  \right\rangle
\right\vert ^{2}/\bar{n}$ versus $t_{h}/t_{2}$ assuming no inhomogeneities,
$\bar{n}=2.5,$ and the harmonic oscillator value $\beta=1.34....$ With
$\xi=0.07$, $U_{3}=0,$ and $\omega/2\pi=30$ kHz, the effective three-body
frequency is $\tilde{U}_{3}/h\simeq-200$ Hz, and $\tilde{U}_{2}/h\simeq2.1$
kHz. The relatively small effective three-body interactions have a strong
effect on the coherence of the state and the resulting quantum interference,
showing how collapse and revival measurements can be a sensitive probe of
coherent higher-body effects. The dephasing is faster than may have been
expected from the small size of $\tilde{U}_{3}$ because the three-body
energies scale as $\tilde{U}_{3}n^{3}$ versus $\tilde{U}_{2}n^{2}$ for
two-body energies, and thus have an increased influence on higher-number
components of a coherent state. Similarly, coherent states with significant
$n>4$ atom number components will probe the four- and higher-body interaction
energies. The bold (red) solid line labeled (iv) in Fig.\ 3 shows the combined
effect of both $\sim$5\% inhomogeneities in $\tilde{U}_{2}$ and three-body
interactions.%

\begin{figure}
[ptb]
\begin{center}
\includegraphics[
height=2.8698in,
width=5.2691in
]%
{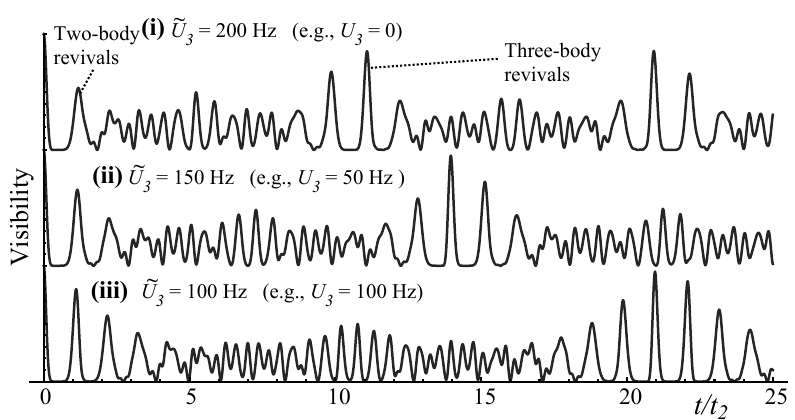}%
\caption{The figure shows the collapse and revival visibility versus time for
$\tilde{U}_{3}=200,150,$ and $50$ Hz assuming negligible inhomogeneities.
Curve (i) corresponds to $U_{3}=0,\xi=0.07,\beta=1.34,\bar{n}=2.5,$ and
$\omega/2\pi=30$ kHz. Curves (ii) and (iii) correspond to smaller three-body
energies $\tilde{U}_{3}$, which could be due to a non-zero intrinsic $U_{3},$
a reduction in $\beta,$ or a change in other system parameters including $\xi$
or $\bar{n}.$ Three-body revivals occur at multiples of $t_{3}=h/\tilde{U}%
_{3},$ providing a method for measuring the coherent three-body interaction
energy. }%
\end{center}
\end{figure}

The decay of the visibility in Fig.~(3) is faster than what is seen in
\cite{Greiner2002 Collapse and Revival,Anderlini2006,JSebbyStrabley1}. Figure
4 illustrates the sensitivity of the evolution of the visibility to the
three-body energy scale by showing three cases corresponding to $\tilde{U}%
_{3}/h=\left\{  -200,-150,-100\right\}  $ Hz. The curves have been displaced
vertically for clarity. Curve (i) for $\tilde{U}_{3}/h=-200$ Hz corresponds to
$U_{3}=0,$ $\beta=1.34$, $\xi=0.07,$ and $\omega/2\pi=30$ kHz. Curve (ii)
corresponds to a reduced $\tilde{U}_{3}/h=-150$ Hz, which could be the result,
for example, of a positive intrinsic three-body energy $U_{3}/h=50$ kHz, or a
change in parameters giving either $\beta\rightarrow3\beta/4$ or
$\xi\rightarrow\sqrt{3}\xi/2$. Similarly, curve (iii) corresponds to
$\tilde{U}_{3}=100$ Hz, which could be due to a positive intrinsic three-body
energy $U_{3}/h=100$ kHz, or to a change in parameters giving either
$\beta\rightarrow\beta/2$ or $\xi\rightarrow\xi/\sqrt{2}.$ The collapse and
revival visibilities are also very sensitive to the average atom number
$\bar{n}.$ A smaller value of $\tilde{U}_{3}$ appears to agree better with the
initial damping seen in \cite{Greiner2002 Collapse and
Revival,Anderlini2006,JSebbyStrabley1}, and this may indicate the presence of
a non-zero intrinsic $U_{3}.$ However, accurate measurement of the system
parameters is necessary if a value of the intrinsic $U_{3}$ is to be obtained
using $U_{3}=\tilde{U}_{3}-\delta U_{3}.$ Nevertheless, it is clear from
Fig.~(4) that both intrinsic and induced three-body interactions can be
important on experimentally relevant timescales.

Figure 4 also shows the partial and full revivals resulting from the beating
between two- and three-body frequency scales expected if inhomogeneities are
sufficiently reduced. The period for nearly full three-body revivals
$t_{3}=h/\tilde{U}_{3}$ gives a direct method of measuring $\tilde{U}_{3}.$
Recently, long sequences of collapse and revivals showing multiple frequencies
have been reported \cite{DAMOP2009}; our analysis suggests that these may be
used to study higher-body interactions in optical lattices.

\section{Summary}

We have shown that two-body induced virtual excitations of bosons to higher
bands in a deep 3D optical lattice generate effective three-body and
higher-body interactions. Although our methods do not yield the intrinsic
higher-body interaction energies $U_{m}$, we find that even if $U_{m}\simeq0$
there are significant effective interactions that can have a surprisingly
strong influence on the dynamics of non-equilibrium coherent states. The
mechanism for higher-body interactions is based upon the recognition that at
low energies the presence of excited (i.e. higher-energy) vibrational states
manifest as $m$-body terms in an effective Hamiltonian $\tilde{H}_{\text{eff}%
}.$ While it is possible for an effective (or renormalized) $m$-body
interaction to vanish or to be very small due to close cancellation of the
intrinsic (i.e. $U_{m}$) and induced (i.e. $\delta U_{m}$) contributions to
$\tilde{U}_{m},$ we do not expect this to happen in general. It is possible to
tune the relative effective $m$-body interactions by exploiting Feshbach
resonances to control $a_{\text{scat}}$, or by changing the lattice potential.
This suggests intriguing possibilities for probing and controlling the physics
of effective field theories (e.g., effective interactions, running coupling
constants, and the emergence of non-perturbative effects) in optical lattices.
Using optical lattices to simulate the controlled breakdown of an effective
field theory would be particularly interesting.

We would like to thank J. Sebby-Strabley and W.D. Phillips for very helpful
conversations. P.R.J. acknowledges support from the Research Corporation for
Science Advancement. J.V.P. acknowledges support from IARPA.


\begin{thebibliography}{99}
\bibitem {3Boby recombo footnote}Using the three-body loss rate $\simeq
6\times10^{-30}$ cm$^{6}$/s, the loss rate for $n$ atoms in a $40$ kHz lattice
site is $2.5n^{2}/s,$ and the lifetime for $n=3$ atoms is $\gtrsim60$ ms.
\bibitem {Calculation of U3}An analytic expression for $K_{\mu0}$ can be
obtained by expressing $K_{\mu0}$ as an integral over a generalized Laguerre
polynomial, and then using an integral representation for the Laguerre
polynomial. The sum in $\delta U_{3}$ can then be solved exactly with
$\Lambda\rightarrow\infty.$
\end{thebibliography}
\end{document}